# Metric Tensor Formulation of Strain in Density-Functional Perturbation Theory

D. R. Hamann,[1,2,3] Xifan Wu,[1] Karin M. Rabe,[1] and David Vanderbilt[1]

[1]Department of Physics and Astronomy, Rutgers University, Piscataway, NJ 08854-8019
[2]Bell Laboratories, Lucent Technologies, Murray Hill, NJ, 07974
[3]Mat-Sim Research LLC, P. O. Box 742, Murray Hill, NJ, 07974

## ABSTRACT

The direct calculation of the elastic and piezoelectric tensors of solids can be accomplished by treating homogeneous strain within the framework of density-functional perturbation theory. By formulating the energy functional in reduced coordinates, we show that the strain perturbation enters only through metric tensors, and can be treated in a manner exactly paralleling the treatment of other perturbations. We present an analysis of the strain perturbation of the plane-wave pseudopotential functional, including the internal strain terms necessary to treat the atomic-relaxation contributions. Procedures for computationally verifying these expressions by comparison with numerical derivatives of ground-state calculations are described and illustrated.

PACS numbers: 71.15.-m, 62.20.Dc, 77.65.Bn, 71.15.Mb

## I. INTRODUCTION

Two seminal contributions to the theory of the electronic structure of solids were the quantum mechanical theory of stress[1] and density-functional perturbation theory.[2] The ability to calculate stress was readily incorporated into density-functional pseudopotential calculations of the ground-state total energies of solids, and finite-difference derivatives of the stress with respect to strain deformations of the unit cell were shown to yield the elastic tensor.[3] Density-functional perturbation theory (DFPT) was widely applied to the direct calculation of phonon spectra, interatomic force constants, Born effective charges, dielectric tensors, and a variety of other properties.[4]

The general structure of DFPT is based upon the systematic expansion of the variational expression for the density-functional theory (DFT) total energy[5] in powers of a parameter $\lambda$ characterizing some dependence of the energy functional.[6] Such parameters as the internal atomic coordinates and the macroscopic electric field[7] could be handled in this framework in a conceptually straightforward manner.[8,9] Treating macroscopic strain as a parameter within this formalism, however, was apparently less straightforward. A canonical-transformation approach to this problem introduced by Baroni *et al.*[10] will be reviewed in Sec. IIIA.

The current approach is based on an overall formulation of the DFT energy expression in reduced coordinates, which introduces real- and reciprocal-space metric tensors into every term in this expression. This formulation will be introduced in Sec. IIIB, and the treatment of the strain derivatives of each term will be detailed in Secs. IIIC-H. In these subsections, we will specialize to the plane-wave representation and norm-conserving pseudopotentials.[11] The advantage of the metric tensor approach is that it puts strain on an equal footing with other parameters characterizing the energy functional, and provides a straightforward if sometimes tedious procedure for evaluating the strain derivatives. While only the first and second derivatives necessary for the evaluation of the elastic and piezoelectric tensors within DFPT are presented here, extensions of the formalism to higher derivatives to evaluate such quantities as nonlinear elastic constants and Grüneisen parameters should be straightforward.

The reduced-coordinate metric tensors were previously used by Souza and Martins as dynamical variables in molecular dynamics simulations with variable unit cell shape.[12] This study has some common conceptual elements with the work presented here, but is not related to the utilization of the metric tensors within DFPT. An unrelated use of the real-space metric tensor in DFT was presented by Rogers and Rappe.[13] Their interest was in calculating the stress tensor field as a function which could vary within the unit cell of a periodic system, and could be formulated as a derivative with respect to a Riemannian metric tensor field. This is to be contrasted with the metric tensors treated here, which are constant throughout space, and related to stresses integrated over bounding surfaces of a unit cell.

Sec. II briefly reviews DFPT and introduces notation that will be used subsequently. Sec. III, as indicated above, presents the details of the metric tensor formulation. In Sec. IV we discuss the comparison of the new, DFPT results for elastic and piezoelectric tensors with the old, numerical-derivative approach and present an illustrative example. We discuss both the clamped-atom case in which all the atoms are displaced proportionally to the strain, and the relaxed-atom case, in which only the unit cells are strained and the atomic positions readjust. In Sec. V, we summarize our findings and comment on extensions to other representations of DFT.

## II. DENSITY-FUNCTIONAL PERTURBATION THEORY

We will briefly recap density-functional perturbation theory in its lowest order both for completeness and to point out the differences present in the context of the strain perturbation in the reduced-coordinate formulation. The notation will follow Gonze[8] as closely as possible. The ground-state electronic energy in DFT is derived by minimizing the functional

$$E_{el}\{\psi_\alpha\} = \sum_\alpha^{occ} \langle \psi_\alpha | T + V_{ext} | \psi_\alpha \rangle + E_{Hxc}[n] \qquad (1)$$

subject to the orthogonality constraint $\langle \psi_\alpha | \psi_\beta \rangle = \delta_{\alpha\beta}$ where $T$ is the kinetic energy, $V_{ext}$ the external potential, the sum is over occupied states $\alpha$, and $E_{Hxc}$ is the Hartree and exchange-correlation energy functional of the density

$$n(\mathbf{r}) = \sum_\alpha^{occ} \psi_\alpha^*(\mathbf{r}) \psi_\alpha(\mathbf{r}). \qquad (2)$$

The set of wave functions minimizing $E_{el}$ satisfy the Kohn-Sham equations[5]

$$H|\psi_\alpha\rangle = \varepsilon_\alpha |\psi_\alpha\rangle, \qquad (3)$$

where the Hamiltonian operator is

$$H = T + V_{ext} + \frac{\delta E_{Hxc}}{\delta n} = T + V_{ext} + V_{Hxc}. \qquad (4)$$

Within the framework of the reduced-coordinate formulation, all problems have an invariant unit cell, a cube of unit dimensions, and an invariant basis set, plane waves periodic in this simple cubic lattice. As will be described in detail in Sec. III, the actual cell shape and dimensions are absorbed into the definitions of all the operators acting on this basis set through the introduction of metric tensors in real and reciprocal space. While DFPT is usually formulated as an expansion of the response to changes in $V_{ext}$, in our case the kinetic energy, Hartree energy, and exchange-correlation energy all have explicit strain dependencies, as well as the implicit strain dependence of the latter two through strain-induced changes of the density.

The usual formulation of DFPT posits a dependence of $E_{el}$ on a parameter $\lambda$ and develops $E_{el}(\lambda)$ and all its components in a power series in $\lambda$,[8]

$$X(\lambda) = X^{(0)} + \lambda X^{(1)} + \lambda^2 X^{(2)} + \cdots \qquad (5)$$

where $X$ can be $E_{el}$, $T$, $V_{ext}$, $\psi_\alpha(\mathbf{r})$, $n(\mathbf{r})$, $\varepsilon_\alpha$, or $H$. The lowest-order expansion of the Kohn-Sham equation, Eq. (3), is

$$H^{(0)}|\psi_\alpha^{(0)}\rangle = \varepsilon_\alpha^{(0)}|\psi_\alpha^{(0)}\rangle. \tag{6}$$

The second-order energy $E_{el}^{(2)}$, in a form which is stationary relative to variations in the first order wave functions $\psi^{(1)}$, is a slight generalization of Eq. (13) of Ref. [8],

$$\begin{aligned}
E_{el}^{(2)}\{\psi^{(0)};\psi^{(1)}\} = \sum_\alpha^{occ} &\Big[ \langle\psi_\alpha^{(1)}|(H^{(0)} - \varepsilon_\alpha^{(0)})|\psi_\alpha^{(1)}\rangle + \big(\langle\psi_\alpha^{(1)}|(T^{(1)} + V_{ext}^{(1)})|\psi_\alpha^{(0)}\rangle + \\
&\langle\psi_\alpha^{(0)}|(T^{(1)} + V_{ext}^{(1)})|\psi_\alpha^{(1)}\rangle\big) + \langle\psi_\alpha^{(0)}|(T^{(2)} + V_{ext}^{(2)})|\psi_\alpha^{(0)}\rangle \Big] \\
&+ \frac{1}{2}\iint \frac{\delta^2 E_{Hxc}}{\delta n(\mathbf{r})\delta n(\mathbf{r}')} n^{(1)}(\mathbf{r})n^{(1)}(\mathbf{r}')\,d\mathbf{r}d\mathbf{r}' \\
&+ \int \frac{\partial}{\partial\lambda}\frac{\delta E_{Hxc}}{\delta n(\mathbf{r})}\bigg|_{n^{(0)}} n^{(1)}(\mathbf{r})\,d\mathbf{r} + \frac{1}{2}\frac{\partial^2 E_{Hxc}}{\partial\lambda^2}\bigg|_{n^{(0)}},
\end{aligned} \tag{7}$$

where the first-order density is given by

$$n^{(1)}(\mathbf{r}) = \sum_\alpha^{occ} [\psi_\alpha^{*(1)}(\mathbf{r})\psi_\alpha^{(0)}(\mathbf{r}) + \psi_\alpha^{*(0)}(\mathbf{r})\psi_\alpha^{(1)}(\mathbf{r})]. \tag{8}$$

and $\psi^{(1)}$ is varied subject to the constraint

$$\langle\psi_\alpha^{(0)}|\psi_\beta^{(1)}\rangle = 0 \tag{9}$$

for all occupied states $\alpha$ and $\beta$. In Eq. (7) we have departed from Ref. [8] in representing the $\lambda$ derivatives of $E_{Hxc}$ as partial derivatives to make clear that only the explicit $\lambda$ dependence is to be considered.

The first-order wave functions which minimize $E_{el}^{(2)}$ subject to Eq. (9) satisfy the self-consistent Sternheimer equation[14] which is the Euler-Lagrange equation for this functional,

$$P_c(H^{(0)} - \varepsilon_\alpha^{(0)})P_c|\psi_\alpha^{(1)}\rangle = -P_c H^{(1)}|\psi_\alpha^{(0)}\rangle, \tag{10}$$

where $P_c$ is the projector onto unoccupied states (conduction bands) and

$$\begin{aligned}
H^{(1)} &= T^{(1)} + V_{ext}^{(1)} + V_{Hxc}^{(1)}, \\
V_{Hxc}^{(1)} &= V_{Hxc0}^{(1)} + \int \frac{\delta^2 E_{Hxc}}{\delta n(\mathbf{r})\delta n(\mathbf{r}')} n^{(1)}(\mathbf{r}')\,d\mathbf{r}', \\
V_{Hxc0}^{(1)} &= \frac{\partial}{\partial\lambda}\frac{\delta E_{Hxc}}{\delta n(\mathbf{r})}\bigg|_{n^{(0)}}.
\end{aligned} \tag{11}$$

Eq. (10) can be solved by a variety of methods, including Green's function[2] and conjugate-gradient[8] approaches.

Practical calculations require finite Bloch wave-vector sums to approximate Brillouin-zone (BZ) integrations. In the case of metals, discontinuous changes in state occupancies as eigenvalues at the finite set of **k** points cross the Fermi surface can lead to computational instabilities. Finite-temperature formulations of DFT[15] smooth the variation of occupancy number with eigenvalue and solve this problem. Eq. (10) must be modified in this case for states in a band of energies around the Fermi energy $\varepsilon_F$,[2,16] and the first-order wave functions in this band can be expressed in a form reminiscent of ordinary finite-temperature perturbation theory. While first-order variations of $\varepsilon_F$ vanish for perturbations with finite wave vector, $\varepsilon_F^{(1)}$ and its contributions to $\psi_\alpha^{(1)}$ and hence $n^{(1)}$ must be included for zero-wave-vector perturbations including strain.[2] An expression for $\varepsilon_F^{(1)}$ is given in Eq. (70) of Ref. [2], but we prefer a simple alternative expression,

$$\varepsilon_F^{(1)} = \sum_\alpha \varepsilon_\alpha^{(1)} f_F'(\varepsilon_\alpha^{(0)}) \Big/ \sum_\alpha f_F'(\varepsilon_\alpha^{(0)}), \qquad (12)$$

where $f_F'(\varepsilon)$ is the derivative of the Fermi function (at $\varepsilon_F^{(0)}$ and $kT$), and the first order eigenvalue is given by

$$\varepsilon_\alpha^{(1)} = \langle \psi_\alpha^{(0)} | H^{(1)} | \psi_\alpha^{(0)} \rangle. \qquad (13)$$

We note that the energy dependence of $f_F'$ confines the contributions in the sums in Eq. (12) to states within the band discussed above. Since the self-consistent contributions to $H^{(1)}$ depend on $\varepsilon_F^{(1)}$, it must be converged in the iterative process of solving the Sternheimer equation (as modified for finite $T$).[2,16]

Excepting the diagonal elements of the elastic tensor, all of the quantities we wish to compute involve mixed second derivatives of $E_{el}$ with respect to two different perturbations. The generalization of Eq. (5) to this case is[9]

$$X(\lambda_1, \lambda_2) = X^{(0)} + \lambda_1 X^{(\lambda_1)} + \lambda_2 X^{(\lambda_2)} + \lambda_1 \lambda_2 X^{(\lambda_1 \lambda_2)} + \cdots. \qquad (14)$$

While stationary expressions for such mixed derivatives of $E_{el}$ can be derived, we have in fact implemented these calculations using the simpler non-stationary expression

$$E_{el}^{\lambda_1 \lambda_2} = \sum_\alpha^{occ} \langle \psi_\alpha^{(\lambda_2)} | (T^{(\lambda_1)} + V_{ext}^{(\lambda_1)} + H_{Hxc0}^{(\lambda_1)}) | \psi_\alpha^{(0)} \rangle$$
$$+ \sum_\alpha^{occ} \langle \psi_\alpha^{(0)} | (T^{(\lambda_1 \lambda_2)} + V_{ext}^{(\lambda_1 \lambda_2)}) | \psi_\alpha^{(0)} \rangle + \frac{1}{2} \frac{\partial^2 E_{Hxc}}{\partial \lambda_1 \partial \lambda_2}\bigg|_{n^{(0)}}, \qquad (15)$$

which only requires the first-order wave functions for one of the perturbations.[9] For metals, $\psi_\alpha^{(\lambda_2)}$ derived as discussed in the preceding paragraph can be used, and Fermi weighting factors $f_F(\varepsilon_\alpha^{(0)})$ should be included in the $\alpha$ sums. We will refer to the terms involving only $\psi_\alpha^{(0)}$ and $n^{(0)}$ in Eqs. (7) and (15) as the frozen-wave-function contributions. In the following sections, we will refer to mixed derivatives with respect to a strain component and an internal atomic-coordinate component as "internal strain" (a term whose usage in the literature is somewhat ambiguous).

Calculation of the piezoelectric tensor involves mixed second derivatives of $E_{el}$ with respect to components of the strain $\boldsymbol{\eta}$ and the electric field $\mathcal{E}$. It is beyond the scope of the present discussion to review the modern Berry-phase theory of polarization in solids.[7,17] However, this theory has been successfully applied within DPFT to the mixed derivative with respect to $\mathcal{E}$ and atomic displacements, which yields Born effective charges, among other quantities. A simple alternative expression to Eq. (15) can be derived for that particular case, Eq. (42) of Ref. [9]. The analogous expression for mixed $\eta$-$\mathcal{E}$ derivatives is

$$\frac{\partial^2 E_{el}}{\partial \tilde{\mathcal{E}}_j \partial \eta_{\alpha\beta}} = 2 \frac{\Omega}{(2\pi)^3} \int_{BZ} \sum_m^{occ} \left\langle i\psi_{\mathbf{k}m}^{(\tilde{k}_j)} \middle| \psi_{\mathbf{k}m}^{(\eta_{\alpha\beta})} \right\rangle d\mathbf{k}, \qquad (16)$$

where $\psi^{(k)}$ is the first-order wave function in the presence of the so-called $\partial/\partial k$ perturbation (an intermediate step in computing electric-field perturbed quantities[7,8]), and $\psi^{(\eta)}$ is the first-order wave function for strain. We have replaced the generic $\alpha$ occupied-state subscript by the Bloch wave vector, band pair $\mathbf{k}m$ and explicitly indicated the Brillouin zone (BZ) integration. Our conventions with regard to reduced quantities and vector and tensor components will be explained in the following section. We remark that there are neither frozen-wave-function nor clamped-ion contributions to this mixed derivative.

### III. STRAIN AND INTERNAL STRAIN DERIVATIVES

#### A. Canonical transformation formulation

The application of homogeneous strain to a crystal lattice simply moves the positions of the atoms and hence changes the DFT external potential,[1]

$$V_{ext}(\mathbf{r}) = \sum_{\mathbf{R}} \sum_{\tau}^{cell} V_{\tau}(\mathbf{r} - \tau - \mathbf{R}) \xrightarrow{\eta} V_{ext}^{\eta}(\mathbf{r}) = \sum_{\mathbf{R}} \sum_{\tau}^{cell} V_{\tau}[\mathbf{r} - (1+\eta)\cdot\tau - (1+\eta)\cdot\mathbf{R}], \quad (17)$$

where $\tau$ denotes the positions of atoms within a unit cell, $\mathbf{R}$ is the set of lattice vectors, and $\eta$ is the Cauchy infinitesimal strain tensor.[18] From the point of view of the infinite lattice the difference, $V_{ext}^{\eta} - V_{ext}$, can never be a small perturbation. Within a single unit cell, of course, an infinitesimal strain will produce an infinitesimal change in potential. However, it also changes the boundary conditions, so the perturbed wave functions cannot be expanded in a basis of the unperturbed wave functions, and DFPT is not applicable.

One solution to this problem was proposed by Baroni *et al.*[19] They introduced a *fictitious* strained self-consistent Hamiltonian obtained from the unstrained Hamiltonian through a scale transformation,

$$\tilde{H}_{SCF}^{\eta}(\mathbf{r}, \nabla) = H_{SCF}\left[(1+\eta)^{-1}\cdot\mathbf{r}, (1+\eta)\cdot\nabla\right]. \qquad (18)$$

Eigenfunctions of $\tilde{H}_{SCF}^{\eta}$ obey the same boundary conditions as those of the actual strained Hamiltonian $H_{SCF}^{\eta}$. The spectrum of $\tilde{H}_{SCF}^{\eta}$ is identical to that of the unstrained Hamiltonian since the two are related by a unitary transformation, and the wave functions and charge density $\tilde{n}^{\eta}$ of $\tilde{H}_{SCF}^{\eta}$ are generated by simple transformations of the corresponding unstrained quantities. The energy difference between the fictitious and unstrained systems is easily computed. The strategy is to then compute the energy differ-ence between the system described by $\tilde{H}_{SCF}^{\eta}$ and that described by the real strained Hamiltonian $H_{SCF}^{\eta}$ using DFPT.[19]

One difficulty in carrying this out is that the Hartree and exchange-correlation terms in $\tilde{H}_{SCF}^{\eta}$ are not the Hartree and xc potentials produced by $\tilde{n}^{\eta}$. However, $\tilde{H}_{SCF}^{\eta}$ can be interpreted as a genuine Kohn-Sham Hamiltonian by modifying the external potential.[19]

While we don't question the validity of this two-step approach, it does change the structure of the calculations from that of ordinary, periodicity-preserving perturbations such as changes in internal atomic coordinates $\tau$. Moreover, Baroni *et al.* present their analysis in terms of uniform dilation and local potentials,[19] and the steps to treat arbitrary strains and non-local pseudopotentials appear to be rather non-trivial within their formulation.

Another formulation for the direct calculation of the DFT elastic tensor was given by Hebbache.[20] While citing the work of Baroni *et al.*,[19] this author included only the frozen-wave-function contributions, and failed to consider the $\psi^{(1)}$ and $n^{(1)}$ contributions to $E^{(2)}$ shown in Eq. (7).

## B. Reduced-coordinate formulation

The reduced coordinates are defined in real space using the basis of three primitive lattice vectors $\mathbf{R}_i^P$ ordered according to their index $i$ to form a right-handed coordinate system. We will follow the convention of using Latin indices $i, j, k,\ldots$ running from 1 to 3 to indicate reduced-coordinate components, and Greek indices $\alpha, \beta, \gamma, \ldots$ to indicate Cartesian components.[21] Thus the components of the primitive lattice are $R_{\alpha i}^P$, those of the primitive reciprocal lattice vectors $\mathbf{G}_j^P$ are $G_{\alpha j}^P$, and the pair satisfy the relationship

$$\sum_{\alpha} R_{\alpha i}^P G_{\alpha j}^P = 2\pi \delta_{ij}, \qquad (19)$$

where the summation range 1,3 will be understood for Cartesian and reduced components throughout. We will notate the reduced counterparts of vectors using a tilde, so a real-space vector $\mathbf{X}$ and its counterpart $\tilde{\mathbf{X}}$ are related by

$$X_\alpha = \sum_i R_{\alpha i}^P \tilde{X}_i. \qquad (20)$$

We will denote the sum of a Bloch vector in the first Brillouin zone and a reciprocal lattice vector by $\mathbf{K} = \mathbf{k} + \mathbf{G}$, and the reduced counterpart by $\tilde{\mathbf{K}}$, with components related by

$$K_\alpha = \sum_i G_{\alpha i}^P \tilde{K}_i. \qquad (21)$$

Essentially every term in the electron energy functional can be expressed as dot products of vectors in real or reciprocal space. The introduction of the metric tensors $\Xi$ for real space and $\Upsilon$ for reciprocal space,

$$\Xi_{ij} = \sum_\alpha R^{\text{P}}_{\alpha i} R^{\text{P}}_{\alpha j}, \quad \Upsilon_{ij} = \sum_\alpha G^{\text{P}}_{\alpha i} G^{\text{P}}_{\alpha j} \tag{22}$$

allows us to express dot products (in real units) in terms of reduced vector components, for example

$$\mathbf{K}' \cdot \mathbf{K} = \sum_{ij} \tilde{K}'_i \Upsilon_{ij} \tilde{K}_j. \tag{23}$$

One further quantity that enters into the energy functional, the unit cell volume $\Omega$, can also be expressed in terms of either metric tensor, for example as $(\det[\Xi_{ij}])^{1/2}$, but the special dependence of $\Omega$ on strain leads us to represent it as a separate entity.

The advantage we obtain from formulating DFT in reduced coordinates is that the boundary conditions never change. The unit cell is a unit cube. Granted, the price we pay for this is a pervasive dependence of all the components of the reduced-coordinate self-consistent Hamiltonian on strain through the metric tensors. However, these are all straightforward parametric dependencies, similar in every way to dependencies on parameters such as internal atomic coordinates, and DFPT can be applied in a straightforward manner. We will derive expressions for the various terms entering into $H^{(1)}$, $H^{(2)}$, and other components of the 2$^{\text{nd}}$-order energy in Secs. IIIC to IIIH below.

The derivatives of real space and reciprocal space vectors with respect to strain are[3]

$$\frac{\partial X_\gamma}{\partial \eta_{\alpha\beta}} = \delta_{\alpha\gamma} X_\beta, \quad \frac{\partial K_\gamma}{\partial \eta_{\alpha\beta}} = -\delta_{\alpha\gamma} K_\beta. \tag{24}$$

Applying these rules to the metric tensors, we find that their first and second strain derivatives are

$$\Xi^{(\alpha\beta)}_{ij} \equiv \frac{\partial \Xi_{ij}}{\partial \eta_{\alpha\beta}} = R^{\text{P}}_{\alpha i} R^{\text{P}}_{\beta j} + R^{\text{P}}_{\beta i} R^{\text{P}}_{\alpha j}, \tag{25}$$

$$\Upsilon^{(\alpha\beta)}_{ij} \equiv \frac{\partial \Upsilon_{ij}}{\partial \eta_{\alpha\beta}} = -G^{\text{P}}_{\alpha i} G^{\text{P}}_{\beta j} - G^{\text{P}}_{\beta i} G^{\text{P}}_{\alpha j}, \tag{26}$$

and

$$\Xi^{(\alpha\beta\gamma\delta)}_{ij} \equiv \frac{\partial^2 \Xi_{ij}}{\partial \eta_{\gamma\delta} \partial \eta_{\alpha\beta}} = \delta_{\alpha\gamma}(R^{\text{P}}_{\beta i} R^{\text{P}}_{\delta j} + R^{\text{P}}_{\delta i} R^{\text{P}}_{\beta j}) + \delta_{\beta\gamma}(R^{\text{P}}_{\alpha i} R^{\text{P}}_{\delta j} + R^{\text{P}}_{\delta i} R^{\text{P}}_{\alpha j}) \\ + \delta_{\alpha\delta}(R^{\text{P}}_{\beta i} R^{\text{P}}_{\gamma j} + R^{\text{P}}_{\gamma i} R^{\text{P}}_{\beta j}) + \delta_{\beta\delta}(R^{\text{P}}_{\alpha i} R^{\text{P}}_{\gamma j} + R^{\text{P}}_{\gamma i} R^{\text{P}}_{\alpha j}), \tag{27}$$

$$\Upsilon_{ij}^{(\alpha\beta\gamma\delta)} \equiv \frac{\partial^2 \Upsilon_{ij}}{\partial \eta_{\gamma\delta} \partial \eta_{\alpha\beta}} = \delta_{\alpha\gamma}(G^{\text{P}}_{\beta i} G^{\text{P}}_{\delta j} + G^{\text{P}}_{\delta i} G^{\text{P}}_{\beta j}) + \delta_{\beta\gamma}(G^{\text{P}}_{\alpha i} G^{\text{P}}_{\delta j} + G^{\text{P}}_{\delta i} G^{\text{P}}_{\alpha j})$$
$$+ \delta_{\alpha\delta}(G^{\text{P}}_{\beta i} G^{\text{P}}_{\gamma j} + G^{\text{P}}_{\gamma i} G^{\text{P}}_{\beta j}) + \delta_{\beta\delta}(G^{\text{P}}_{\alpha i} G^{\text{P}}_{\gamma j} + G^{\text{P}}_{\gamma i} G^{\text{P}}_{\alpha j}), \quad (28)$$

where we have introduced the notation of parenthesized Cartesian superscripts to denote strain derivatives. It can be verified that these formulas are invariant under interchange of $(\alpha, \beta)$ or $(\gamma, \delta)$ index pairs. This is a manifestation of the fact that antisymmetric components of $\boldsymbol{\eta}$ correspond to rotations rather than strains, under which the metric tensors are invariant.

The strain derivative of the unit-cell volume $\Omega$ is sufficiently simple so as not to warrant additional notation,

$$\frac{\partial \Omega}{\partial \eta_{\alpha\beta}} = \delta_{\alpha\beta} \Omega. \quad (29)$$

The extension to second derivatives is obvious. Finally, it is easily shown from Eq. (19) that

$$\mathbf{K} \cdot \mathbf{X} = 2\pi \tilde{\mathbf{K}} \cdot \tilde{\mathbf{X}}, \quad (30)$$

so dot products between real and reciprocal vectors do not involve the metric tensors and are strain independent.

We note that DFPT yields second derivatives of the energy per unit cell. This has the consequence that the naturally defined "elastic tensor" as calculated in DFPT,

$$C^*_{\alpha\beta\gamma\delta} \equiv \frac{1}{\Omega} \frac{\partial^2 E_{el}}{\partial \eta_{\alpha\beta} \partial \eta_{\gamma\delta}}, \quad (31)$$

is not equal to the conventional elastic tensor

$$C_{\alpha\beta\gamma\delta} \equiv \frac{\partial \sigma_{\gamma\delta}}{\partial \eta_{\alpha\beta}} = \frac{\partial}{\partial \eta_{\alpha\beta}} \frac{1}{\Omega} \frac{\partial E_{el}}{\partial \eta_{\gamma\delta}} = C^*_{\alpha\beta\gamma\delta} - \delta_{\alpha\beta} \sigma_{\gamma\delta}, \quad (32)$$

where $\sigma_{\gamma\delta}$ is the stress tensor. If the reference state of the system has had its lattice parameters fully relaxed, $\mathbf{C}^*$ and $\mathbf{C}$ are identical. However, for calculations of the elastic tensor of materials under stress, Eq. (32) gives important corrections, and the Voigt symmetry under the interchange $\alpha\beta \leftrightarrow \gamma\delta$ can be violated.[22]

Finally, we point out that when higher-order elastic properties are to be considered as extensions of this approach, the connection between the Cauchy infinitesimal strain and the conventional Lagrangian strain needs to be taken into account.[1,18]

## C. Kinetic energy

The wave functions $\psi^{(0)}_\alpha$ and $\psi^{(1)}_\alpha$ are to be expanded as sums of reduced plane waves,

$$\left| \psi_{\tilde{\mathbf{k}}\alpha} \right\rangle = \sum_{\tilde{\mathbf{G}}} c_{\tilde{\mathbf{k}}\alpha\tilde{\mathbf{G}}} \left| \tilde{\mathbf{K}} \right\rangle, \quad (33)$$

so most of the operators involved in the Sternheimer equation and the second-order energies will be expressed in terms of their reduced plane-wave matrix elements. The strain derivatives of the kinetic energy, which remains a diagonal operator in the reduced plane-wave basis, are rather trivially found from the metric tensor derivatives given in the previous section. However, in procedures in which the real unit cell varies, such as constant-pressure molecular dynamics or lattice parameter optimization, it may be desirable to add a function $f_{SM}(\varepsilon_{\tilde{K}})$ to the kinetic energy $\varepsilon_{\tilde{K}}$ which smoothly becomes large approaching the plane-wave cutoff energy. This will force the wave-function coefficients to zero at the cutoff and regularize the variation of the energy.[23] While the DFPT calculation is of course done with a fixed unit cell, it may be desirable to keep the smoothing function used in optimizing the cell parameters to ensure that stresses remain below the limit achieved in the optimization. Incorporating this generalization, the reduced-coordinate operators are

$$\langle \tilde{\mathbf{K}}'|T|\tilde{\mathbf{K}}\rangle = [\varepsilon_{\tilde{\mathbf{K}}} + f_{SM}(\varepsilon_{\tilde{\mathbf{K}}})]\delta_{\tilde{\mathbf{K}}'\tilde{\mathbf{K}}}, \quad (34)$$

where

$$\varepsilon_{\tilde{\mathbf{K}}} = \tfrac{1}{2}\sum_{ij}\Upsilon_{ij}\tilde{K}_i\tilde{K}_j, \quad (35)$$

$$\langle \tilde{\mathbf{K}}'|\frac{\partial T}{\partial \eta_{\alpha\beta}}|\tilde{\mathbf{K}}\rangle = \frac{1}{2}\left\{[1+f'_{SM}(\varepsilon_{\tilde{\mathbf{K}}})]\sum_{ij}\Upsilon^{(\alpha\beta)}_{ij}\tilde{K}_i\tilde{K}_j\right\}\delta_{\tilde{\mathbf{K}}'\tilde{\mathbf{K}}}, \quad (36)$$

$$\langle \tilde{\mathbf{K}}'|\frac{\partial^2 T}{\partial \eta_{\alpha\beta}\partial \eta_{\gamma\delta}}|\tilde{\mathbf{K}}\rangle = \left\{f''_{SM}(\varepsilon_{\tilde{\mathbf{K}}})\left[\frac{1}{2}\sum_{ij}\Upsilon^{(\alpha\beta)}_{ij}\tilde{K}_i\tilde{K}_j\right]^2 \right.$$
$$\left. +\frac{1}{2}[1+f'_{SM}(\varepsilon_{\tilde{\mathbf{K}}})]\sum_{ij}\Upsilon^{(\alpha\beta\gamma\delta)}_{ij}\tilde{K}_i\tilde{K}_j\right\}\delta_{\tilde{\mathbf{K}}'\tilde{\mathbf{K}}}, \quad (37)$$

and primes denote derivatives of $f_{SM}$. The kinetic energy operator has no explicit dependence on atomic positions, so the mixed second derivative term for internal strain is zero.

### D. Local pseudopotential

Operations of the local pseudopotential component of $V_{ext}$, $V_{Loc}$, on the wave functions are most efficiently evaluated in reduced real space, followed by Fourier transformation to obtain the the $\langle\tilde{\mathbf{K}}|$ components. This applies to the first-order local potential as well, so the strain derivative of $V_{Loc}$ operating on $|\psi^{(0)}_{\tilde{\mathbf{k}}\alpha}\rangle$ is evaluated as

$$\langle\tilde{\mathbf{K}}|\frac{\partial V_{Loc}}{\partial \eta_{\alpha\beta}}|\psi^{(0)}_{\tilde{\mathbf{k}}\alpha}\rangle = \int e^{-2\pi i \tilde{\mathbf{K}}\cdot\tilde{\mathbf{r}}} V^{(\alpha\beta)}_{Loc}(\tilde{\mathbf{r}})\psi^{(0)}_{\tilde{\mathbf{k}}\alpha}(\tilde{\mathbf{r}}) d^3\tilde{r}. \quad (38)$$

The first-order potential itself is most conveniently evaluated in reciprocal space. Following Eq. (23), the squared magnitude of the reciprocal lattice vectors expressed in terms of reduced coordinates is

$$G^2 = \sum_{ij} \Upsilon_{ij} \tilde{G}_i \tilde{G}_j. \tag{39}$$

The potential components are given by

$$V_{Loc}^{(\alpha\beta)}(\tilde{\mathbf{G}}) = \frac{1}{\Omega} \sum_{\kappa}^{cell} e^{-2\pi i \tilde{\mathbf{G}} \cdot \tilde{\tau}_\kappa} \left[ -\delta_{\alpha\beta} v_{\kappa Loc}(G) + \frac{v'_{\kappa Loc}(G)}{2G} \sum_{ij} \Upsilon_{ij}^{(\alpha\beta)} \tilde{G}_i \tilde{G}_j \right], \tag{40}$$

where $v_{\kappa Loc}$ is the Fourier transform of the local pseudopotential of the atom $\kappa$ at site $\tilde{\tau}_\kappa$,

$$v_{\kappa Loc}(G) = 4\pi \int_0^\infty j_0(Gr) v_{\kappa Loc}(r) r^2 dr, \tag{41}$$

and $v'_{\kappa Loc}$ is its first derivative. We have omitted the conventional $\Omega^{-1}$ normalization in Eq. (41) and placed it in Eq. (40) so that the Fourier transform atomic potentials depend on strain only through their arguments. We note that the phases (or structure factors) do not contribute to the strain derivatives.

The second derivative of the local pseudopotential energy with respect to two strains occurring in Eq. (15) can be expressed entirely in terms of the Fourier components of the zero-order density,

$$\begin{aligned}
\frac{\partial^2 E_{Loc}}{\partial \eta_{\alpha\beta} \partial \eta_{\gamma\delta}} = \sum_{\tilde{\mathbf{G}} \neq 0} n_{\tilde{\mathbf{G}}}^{(0)} \sum_{\kappa}^{cell} e^{-2\pi i \tilde{\mathbf{G}} \cdot \tilde{\tau}_\kappa} &\left[ \delta_{\alpha\beta} \delta_{\gamma\delta} v_{\kappa Loc}(G) - \frac{v'_{\kappa Loc}(G)}{2G} \sum_{ij} \left( \delta_{\alpha\beta} \Upsilon_{ij}^{(\gamma\delta)} + \delta_{\gamma\delta} \Upsilon_{ij}^{(\alpha\beta)} \right. \right. \\
&\left. \left. - \Upsilon_{ij}^{(\alpha\beta\gamma\delta)} \right) \tilde{G}_i \tilde{G}_j + \left( \frac{v''_{\kappa Loc}(G)}{4G^2} - \frac{v'_{\kappa Loc}(G)}{4G^3} \right) \sum_{ij} \Upsilon_{ij}^{(\alpha\beta)} \tilde{G}_i \tilde{G}_j \sum_{kl} \Upsilon_{kl}^{(\gamma\delta)} \tilde{G}_k \tilde{G}_l \right],
\end{aligned} \tag{42}$$

where $v''_{\kappa Loc}$ is the second derivative. Finally, mixed second derivatives with respect to one strain component and one reduced-atomic-coordinate component are required for internal strain,

$$\frac{\partial^2 E_{Loc}}{\partial \eta_{\alpha\beta} \partial \tilde{\tau}_{\kappa k}} = -2\pi i \sum_{\tilde{\mathbf{G}}} n_{\tilde{\mathbf{G}}}^{(0)} \tilde{G}_k e^{-2\pi i \tilde{\mathbf{G}} \cdot \tilde{\tau}_\kappa} \left[ -\delta_{\alpha\beta} v_{\kappa Loc}(G) + \frac{v'_{\kappa Loc}(G)}{2G} \sum_{ij} \Upsilon_{ij}^{(\alpha\beta)} \tilde{G}_i \tilde{G}_j \right]. \tag{43}$$

### E. Non-local pseudopotential

The first strain derivative of the semi-local form of norm-conserving pseudopotentials[11] was given by Nielsen and Martin.[3] The fully separable form introduced by Kleinman and Bylander[24] and its generalization by Blöchl[25] are far more widely used today because of their computational efficiency. The matrix elements of the nonlocal pseudopotentials are most commonly expressed in the form

$$\langle \mathbf{K}' | V_{NL} | \mathbf{K} \rangle = \frac{1}{\Omega} \sum_{\kappa \ell m} e^{i\mathbf{K}' \cdot \tau_\kappa} v_{\kappa\ell}(|\mathbf{K}'|) Y_{\ell m}(\theta_{\mathbf{K}'}, \phi_{\mathbf{K}'}) e^{-i\mathbf{K} \cdot \tau_\kappa} v_{\kappa\ell}(|\mathbf{K}|) Y_{\ell m}(\theta_{\mathbf{K}}, \phi_{\mathbf{K}}) \tag{44}$$

where each Fourier-transformed separable atomic potential is

$$v_{\kappa\ell}(|\mathbf{K}|) = 4\pi \int_0^\infty j_\ell(|\mathbf{K}|r) v_{\kappa\ell}(r) r^2 dr, \qquad (45)$$

$v_{\kappa\ell}(r)$ is the real-space potential in angular momentum channel $\ell$ for the atom $\kappa$, and $j_\ell$ are spherical Bessel functions. We show the single-projector form, but the generalization to more projectors[25] is obvious. We have omitted the conventional $\Omega^{-1/2}$ in Eq. (45) as in the local case in Sec. D. The first strain derivative of Eq. (44) was initially given by Bylander et al.,[26] but their expression had substantial omissions which were corrected by I.-H. Lee et al.[27] The resulting expression is quite cumbersome, not suitable for evaluation in terms of reduced coordinates and the metric tensors, and appears to be extremely difficult to extend to higher derivatives.

To transform Eq. (44) so that it is suitable for our purposes, we explicitly carry out the $m$ sum to obtain

$$\langle \mathbf{K}' | V_{NL} | \mathbf{K} \rangle = \frac{4\pi}{\Omega} \sum_{\kappa\ell} (2\ell+1) e^{i\mathbf{K}'\cdot\boldsymbol{\tau}_\kappa} v_{\tau\ell}(|\mathbf{K}'|) P_\ell(\cos\theta_{\mathbf{K}',\mathbf{K}}) e^{-i\mathbf{K}\cdot\boldsymbol{\tau}_\kappa} v_{\kappa\ell}(|\mathbf{K}|), \qquad (46)$$

where $P_\ell$ are Legendre polynomials and $\theta_{\mathbf{K}',\mathbf{K}}$ is the angle between $\mathbf{K}'$ and $\mathbf{K}$. Introducing the modified function

$$\wp_\ell(\mathbf{K}'\cdot\mathbf{K}', \mathbf{K}'\cdot\mathbf{K}, \mathbf{K}\cdot\mathbf{K}) \equiv 4\pi(2\ell+1) |\mathbf{K}'|^\ell |\mathbf{K}|^\ell P_\ell(\cos\theta_{\mathbf{K}',\mathbf{K}}), \qquad (47)$$

where $\wp_\ell$ is a polynomial in the three dot products, and the modified potential form factor

$$f_{\kappa\ell}(\mathbf{K}\cdot\mathbf{K}) \equiv v_{\kappa\ell}(|\mathbf{K}|)/|\mathbf{K}|^\ell, \qquad (48)$$

we reformulate Eq. (46) as

$$\langle \mathbf{K}' | V_{NL} | \mathbf{K} \rangle = \frac{1}{\Omega} \sum_{\kappa\ell m} e^{i\mathbf{K}'\cdot\boldsymbol{\tau}_\kappa} f_{\kappa\ell}(\mathbf{K}'\cdot\mathbf{K}') \wp_\ell(\mathbf{K}'\cdot\mathbf{K}', \mathbf{K}'\cdot\mathbf{K}, \mathbf{K}\cdot\mathbf{K}) e^{-i\mathbf{K}\cdot\boldsymbol{\tau}_\kappa} f_{\kappa\ell}(\mathbf{K}\cdot\mathbf{K}). \qquad (49)$$

Eq. (49) is now straightforward to express in reduced coordinates. First, we observe that the phases constituting the structure factors are independent of the metric tensors, $\mathbf{K}\cdot\boldsymbol{\tau}_\kappa = 2\pi\tilde{\mathbf{K}}\cdot\tilde{\boldsymbol{\tau}}_\kappa$, and will thus be independent of strain. After introducing the metric tensors and reduced wave vectors in $\wp_\ell$, we obtain

$$\langle \tilde{\mathbf{K}}' | V_{NL} | \tilde{\mathbf{K}} \rangle = \frac{1}{\Omega} \sum_{\kappa\ell} e^{2\pi i \tilde{\mathbf{K}}'\cdot\tilde{\boldsymbol{\tau}}_\kappa} f_{\kappa\ell}(\sum_{ij} \Upsilon_{ij} \tilde{K}'_i \tilde{K}'_j) \times$$
$$\wp_\ell(\sum_{ij} \Upsilon_{ij} \tilde{K}'_i \tilde{K}'_j, \sum_{ij} \Upsilon_{ij} \tilde{K}'_i \tilde{K}_j, \sum_{ij} \Upsilon_{ij} \tilde{K}_i \tilde{K}_j) e^{-2\pi i \tilde{\mathbf{K}}\cdot\tilde{\boldsymbol{\tau}}_\kappa} f_{\kappa\ell}(\sum_{ij} \Upsilon_{ij} \tilde{K}_i \tilde{K}_j) \qquad (50)$$

If $\wp_\ell$ is expanded, we observe that it is a polynomial in which all terms are products of $\ell$ components $\tilde{K}'_i$ and $\ell$ components $\tilde{K}_i$. We can regroup terms and formulate Eq. (50) in terms of such tensor products,[28]

$$T_{\ell m}(\tilde{\mathbf{K}}) = \tilde{K}_1^{I_T(1,\ell,m)} \tilde{K}_2^{I_T(2,\ell,m)} \tilde{K}_3^{\ell - I_T(1,\ell,m) - I_T(2,\ell,m)} \tag{51}$$

where $I_T(i,\ell,m)$ is an indexing array of non-negative integers. This array can be defined in a systematic way for tensors from rank 0 up to the highest we shall encounter. The $m$ index runs from 1 to $(\ell+1)(\ell+2)/2$. The matrix element can then be expressed as

$$\langle \tilde{\mathbf{K}}' | V_{NL} | \tilde{\mathbf{K}} \rangle = \frac{1}{\Omega} \sum_{\kappa \ell m m'} e^{2\pi i \tilde{\mathbf{K}}' \cdot \tilde{\boldsymbol{\tau}}_\kappa} f_{\kappa \ell}(\sum_{ij} \Upsilon_{ij} \tilde{K}_i' \tilde{K}_j') T_{\ell m'}(\tilde{\mathbf{K}}') C_{\ell m' m}(\Upsilon_{ij}) \times$$
$$e^{-2\pi i \tilde{\mathbf{K}} \cdot \tilde{\boldsymbol{\tau}}_\kappa} f_{\kappa \ell}(\sum_{ij} \Upsilon_{ij} \tilde{K}_i \tilde{K}_j) T_{\ell m}(\tilde{\mathbf{K}}), \tag{52}$$

where each $C_{\ell m m'}$ is a polynomial in the components of $\Upsilon$, whose coefficients can be calculated once for all.[28] The notation in Eq. (52) has been chosen to resemble that of Eq. (44), so that its fully separable form is clear. However, the $m$ and $m'$ terms are coupled both because the $T_{\ell m}$ tensors do not form an orthogonal set like the $Y_{\ell m}$, and because the shapes of the angular projectors are no longer spherical harmonics when mapped into reduced coordinates. There is no coupling among different angular momenta $\ell$, however, because deformations cannot change the number of nodes of the projectors.

The procedure for evaluating strain derivatives is now completely straightforward. The operator $\partial/\partial \eta_{\alpha\beta}$ applied to Eq. (50) will act on the $\Omega^{-1}$ prefactor, on the $\Upsilon_{ij}$ coefficients in the $\wp_\ell$ polynomial, and on the arguments of the $f_{\kappa\ell}$. Defining the $v^{\text{th}}$ derivative of $f_{\tau \ell}$ with respect to its argument as $f_{\tau \ell}^{(v)}$ where $v = 0, 1, 2, \ldots$, we observe

$$\frac{\partial f_{\kappa\ell}^{(0)}}{\partial \eta_{\alpha\beta}} = f_{\kappa\ell}^{(1)} \sum_{ij} \Upsilon_{ij}^{(\alpha\beta)} \tilde{K}_i \tilde{K}_j, \tag{53}$$

so this derivative raises the rank of one of the tensor products by 2. The derivative of Eq. (50) can be written in a form very similar to Eq. (52),

$$\langle \tilde{\mathbf{K}}' | \frac{\partial V_{NL}}{\partial \eta_{\alpha\beta}} | \tilde{\mathbf{K}} \rangle = \frac{1}{\Omega} \sum_{\kappa \ell m m' v' v} e^{2\pi i \tilde{\mathbf{K}}' \cdot \tilde{\boldsymbol{\tau}}_\kappa} f_{\kappa\ell}^{(v')}(\sum_{ij} \Upsilon_{ij} \tilde{K}_i' \tilde{K}_j') T_{\ell+2v',m'}(\tilde{\mathbf{K}}') \times$$
$$C_{\ell m' m v' v}^{\alpha\beta}(\Upsilon_{ij}, \Upsilon_{ij}^{(\alpha\beta)}) e^{-2\pi i \tilde{\mathbf{K}} \cdot \tilde{\boldsymbol{\tau}}_\kappa} f_{\kappa\ell}^{(v)}(\sum_{ij} \Upsilon_{ij} \tilde{K}_i \tilde{K}_j) T_{\ell+2v,m}(\tilde{\mathbf{K}}) \tag{54}$$
$$-\delta_{\alpha\beta} \langle \tilde{\mathbf{K}}' | V_{NL} | \tilde{\mathbf{K}} \rangle,$$

where the indices $v, v'$ run from 0 to 1 subject to $v + v' \leq 1$, the $m$ index runs from 1 to $(\ell+2v+1)(\ell+2v+2)/2$, and similarly for $m'$ (with $v \to v'$). The $C_{\ell m' m v' v}^{\alpha\beta}$ matrix elements are each polynomials in $\Upsilon_{ij}$ and $\Upsilon_{ij}^{(\alpha\beta)}$. The couplings here can be translated back to more familiar angular momentum terms, since the leading (rank) index of the $T_{\ell m}$ tensors does correspond to the ordinary $\ell$. This derivative operator couples components $\ell$ on the right to $\ell-2$, $\ell$, and $\ell+2$ on the left. The last term arises from the derivative of the $\Omega^{-1}$ prefactor in Eq. (52).

The extension to second strain derivatives, needed in the $<\psi^{(0)}|H^{(2)}|\psi^{(0)}>$ contribution to the second-order energies, is similarly straightforward and can be expressed in nearly the same form,

$$\langle \tilde{\mathbf{K}}'|\frac{\partial^2 V_{NL}}{\partial \eta_{\alpha\beta}\partial \eta_{\gamma\delta}}|\tilde{\mathbf{K}}\rangle = \frac{1}{\Omega}\sum_{\kappa\ell mm'\nu'\nu} e^{2\pi i \tilde{\mathbf{K}}'\cdot\tilde{\tau}_\kappa} f_{\tau\ell}^{(\nu')}(\sum_{ij}\Upsilon_{ij}\tilde{K}'_i\tilde{K}'_j) T_{\ell+2\nu',m'}(\tilde{\mathbf{K}}') \times$$

$$C_{\ell m'm\nu'\nu}^{\alpha\beta\gamma\delta}(\Upsilon_{ij},\Upsilon_{ij}^{(\alpha\beta)},\Upsilon_{ij}^{(\gamma\delta)},\Upsilon_{ij}^{(\alpha\beta\gamma\delta)}) e^{-2\pi i \tilde{\mathbf{K}}\cdot\tilde{\tau}_\kappa} f_{\kappa\ell}^{(\nu)}(\sum_{ij}\Upsilon_{ij}\tilde{K}_i\tilde{K}_j) T_{\ell+2\nu,m}(\tilde{\mathbf{K}}) \quad (55)$$

$$-\delta_{\alpha\beta}\langle \tilde{\mathbf{K}}'|\frac{\partial V_{NL}}{\partial \eta_{\gamma\delta}}|\tilde{\mathbf{K}}\rangle - \delta_{\gamma\delta}\langle \tilde{\mathbf{K}}'|\frac{\partial V_{NL}}{\partial \eta_{\alpha\beta}}|\tilde{\mathbf{K}}\rangle + \delta_{\alpha\beta}\delta_{\gamma\delta}\langle \tilde{\mathbf{K}}'|V_{NL}|\tilde{\mathbf{K}}\rangle,$$

where the indices $\nu,\nu'$ now run from 0 to 2 subject to $\nu+\nu'\leq 2$, the $m,m'$ ranges depend on $\nu,\nu'$ as above, and the $C$ matrix elements are polynomials in components of all the indicated arguments. Here, possible right-to-left angular momentum couplings are $\ell$ to $\ell-4$, $\ell-2$, $\ell$, $\ell+2$, and $\ell+4$.

Finally, we need to consider mixed derivatives with respect to one strain component and one atomic displacement. Differentiating Eq. (50) with respect to the reduced coordinate $\tilde{\tau}_{\kappa k}$ will introduce factors $-2\pi i \tilde{K}'_k$ or $2\pi i \tilde{K}_k$, and our result will be of the form

$$\langle \tilde{\mathbf{K}}'|\frac{\partial^2 V_{NL}}{\partial \tilde{\tau}_{\kappa k}\partial \eta_{\alpha\beta}}|\tilde{\mathbf{K}}\rangle = \frac{2\pi i}{\Omega}\sum_{\substack{\ell mm'\\ \nu\nu'\mu\mu'}} e^{2\pi i \tilde{\mathbf{K}}'\cdot\tilde{\tau}_\kappa} f_{\kappa\ell}^{(\nu')}(\sum_{ij}\Upsilon_{ij}\tilde{K}'_i\tilde{K}'_j) T_{\ell+2\nu'+\mu',m'}(\tilde{\mathbf{K}}') \times$$

$$C_{\ell m'm\nu\nu'\mu\mu'}^{k\alpha\beta}(\Upsilon_{ij},\Upsilon_{ij}^{(\alpha\beta)}) e^{-2\pi i \tilde{\mathbf{K}}\cdot\tilde{\tau}_\kappa} f_{\tau\ell}^{(\nu)}(\sum_{ij}\Upsilon_{ij}\tilde{K}_i\tilde{K}_j) T_{\ell+2\nu+\mu,m}(\tilde{\mathbf{K}}) \quad (56)$$

$$-\delta_{\alpha\beta}\langle \tilde{\mathbf{K}}'|\frac{\partial V_{NL}}{\partial \tilde{\tau}_{\kappa k}}|\tilde{\mathbf{K}}\rangle,$$

where the indices $\nu,\nu'$ run from 0 to 1 subject to $\nu+\nu'\leq 1$, the new index pair $\mu,\mu'$ run from 0 to 1 subject to $\mu+\mu'=1$, and the $m,m'$ indices span the ranges indicated by the rank of the respective $T$ tensors. Here, the angular momentum couplings are $\ell$ to $\ell-3$, $\ell-1$, $\ell+1$, and $\ell+3$. The expression for the atomic-displacement derivative in the last term can be found in Ref. [8].

The task of carrying out the differentiations, collecting terms, and extracting the coefficients of the $T$ tensors to obtain the $C$ matrix element polynomials in Eqs.(52) and (54) through (56) appears to be extremely tedious. However, the structure of this procedure is sufficiently simple that it is easily automated using a symbolic manipulation program.[29] Since they depend only on the primitive lattice vectors, these polynomials need only be evaluated once, and the task of applying the derivative nonlocal potentials to a set of wave functions is computationally comparable to that of applying the potentials themselves. For expectation values such as $<\psi^{(0)}|H^{(2)}|\psi^{(0)}>$, certain pair of $\nu,\nu'$ and $\mu,\mu'$ indices give hermitian conjugate contributions, and the sums over these indices may be simplified accordingly.

### F. Hartree Potential

The operation of the first-order Hartree potential on the zero-order wave functions is evaluated in real space using an analogous expression to that for the local potential, Eq. (38). The potential is most easily calculated in reciprocal space, however, where the Poisson equation is diagonal. The zero-order electron density components $n_{\mathbf{G}}^{(0)}$ depend on strain only through their $\Omega^{-1}$ normalization factor.[1] The Fourier components of the first-order Hartree potential are

$$V_{H\tilde{\mathbf{G}}}^{(\alpha\beta)} = \frac{4\pi}{G^2}\left[n_{\tilde{\mathbf{G}}}^{(\alpha\beta)} - n_{\tilde{\mathbf{G}}}^{(0)}\left(\delta_{\alpha\beta} + \frac{1}{G^2}\sum_{ij}\Upsilon_{ij}^{(\alpha\beta)}\tilde{G}_i\tilde{G}_j\right)\right],\tag{57}$$

where $n_{\tilde{\mathbf{G}}}^{(\alpha\beta)}$ are the Fourier components of the first-order density for the strain perturbation, and $G^2$ is given by Eq. (39). The second-order strain derivatives of the Hartree energy are

$$\frac{\partial^2 E_H}{\partial \eta_{\alpha\beta}\partial\eta_{\gamma\delta}} = 2\pi\Omega\sum_{\tilde{\mathbf{G}}\neq 0} n_{\tilde{\mathbf{G}}}^{(0)*}n_{\tilde{\mathbf{G}}}^{(0)}\left[\delta_{\alpha\beta}\delta_{\gamma\delta}G^{-2} + G^{-4}\sum_{ij}\left(\delta_{\alpha\beta}\Upsilon_{ij}^{(\gamma\delta)} + \delta_{\gamma\delta}\Upsilon_{ij}^{(\alpha\beta)}\right.\right.$$
$$\left.\left. - \Upsilon_{ij}^{(\alpha\beta\gamma\delta)}\right)\tilde{G}_i\tilde{G}_j + 2G^{-6}\sum_{ij}\Upsilon_{ij}^{(\alpha\beta)}\tilde{G}_i\tilde{G}_j\sum_{kl}\Upsilon_{kl}^{(\gamma\delta)}\tilde{G}_k\tilde{G}_l\right],\tag{58}$$

There is no Hartree contribution to the internal strain.

### G. Exchange-correlation potential

The operation of the first-order exchange-correlation potential on the zero-order wave functions is evaluated as in Eq. (38). If the density $n^{(0)}$ consisted only of contributions from the zero-order wave functions, its explicit strain dependence would arise only from the $\Omega^{-1}$ normalization factor, and would be trivially found from Eq. (29).[30] However, it is frequently desirable to include a non-linear core correction through model core charges,[31] which significantly complicates the analysis. In this section, we must distinguish "electron" and "core" contributions, $n^{(0)} = n_e^{(0)} + n_c$, where the core density is given by a sum of finite-range spherically-symmetric atom-centered functions,

$$n_c(\mathbf{r}) = \sum_{\mathbf{R}}\sum_{\kappa}^{\text{cell}}\rho_{\kappa c}\left(|\mathbf{r}-\boldsymbol{\tau}_\kappa-\mathbf{R}|\right).\tag{59}$$

Considering for present purposes only local-density functionals, it is straightforward to show from Eq. (11) that the first-order *xc* potential is

$$V_{xc}^{(\alpha\beta)} \equiv \frac{\partial V_{xc}}{\partial \eta_{\alpha\beta}} = K_{xc}\left(-\delta_{\alpha\beta}n_e^{(0)} + \frac{\partial n_c}{\partial \eta_{\alpha\beta}} + n^{(\alpha\beta)}\right),\tag{60}$$

where we define

$$K_{xc} \equiv \left. \frac{dV_{xc}(n)}{dn} \right|_{n^{(0)}}. \tag{61}$$

We have included in Eq. (60) both the explicit strain dependence of the zero-order densities and the first-order density for the strain perturbation, $n^{(\alpha\beta)}$, which must be evaluated self-consistently through Eqs. (8) and (10). All the terms in Eqs. (60) and (61) are functions of the real or reduced spatial coordinate, these arguments having been omitted for clarity.

The model core charge in reduced coordinates, $n_c(\tilde{\mathbf{r}})$, is a non-trivial function of strain through the arguments of the $\rho_{\kappa c}$. Introducing the notation for the magnitude ("size") of a reduced-coordinate real-space vector

$$s(\tilde{\mathbf{r}}) = \left( \sum_{ij} \Xi_{ij} \tilde{r}_i \tilde{r}_j \right)^{1/2}, \tag{62}$$

and its strain derivative

$$s^{(\alpha\beta)}(\tilde{\mathbf{r}}) \equiv \frac{\partial s(\tilde{\mathbf{r}})}{\partial \eta_{\alpha\beta}} = \frac{1}{2s(\tilde{\mathbf{r}})} \sum_{ij} \Xi_{ij}^{(\alpha\beta)} \tilde{r}_i \tilde{r}_j, \tag{63}$$

we have

$$\frac{\partial n_c(\tilde{\mathbf{r}})}{\partial \eta_{\alpha\beta}} = \sum_{\tilde{\mathbf{R}}}^{\text{cell}} \sum_{\kappa} \rho'_{\kappa c}\left[ s(\tilde{\mathbf{r}} - \tilde{\boldsymbol{\tau}}_\kappa - \tilde{\mathbf{R}}) \right] s^{(\alpha\beta)}(\tilde{\mathbf{r}} - \tilde{\boldsymbol{\tau}}_\kappa - \tilde{\mathbf{R}}), \tag{64}$$

where $\rho'_{\kappa c}$ are the first derivatives of each model core function with respect to its argument.

Second-order $xc$ terms in Eq. (15) for the strain-strain derivatives require corresponding derivatives of the "size" function,

$$s^{(\alpha\beta\gamma\delta)}(\tilde{\mathbf{r}}) \equiv \frac{\partial^2 s(\tilde{\mathbf{r}})}{\partial \eta_{\alpha\beta} \partial \eta_{\gamma\delta}} = -\frac{1}{4s^3(\tilde{\mathbf{r}})} \sum_{ij} \Xi_{ij}^{(\alpha\beta)} \tilde{r}_i \tilde{r}_j \sum_{k\ell} \Xi_{k\ell}^{(\gamma\delta)} \tilde{r}_k \tilde{r}_\ell + \frac{1}{2s(\tilde{\mathbf{r}})} \sum_{ij} \Xi_{ij}^{(\alpha\beta\gamma\delta)} \tilde{r}_i \tilde{r}_j, \tag{65}$$

in terms of which the core charge second derivatives can be evaluated as

$$\frac{\partial^2 n_c(\tilde{\mathbf{r}})}{\partial \eta_{\alpha\beta} \partial \eta_{\gamma\delta}} = \sum_{\tilde{\mathbf{R}}}^{\text{cell}} \sum_{\kappa} \Big\{ \rho'_{\kappa c}\left[ s(\tilde{\mathbf{r}} - \tilde{\boldsymbol{\tau}}_\kappa - \tilde{\mathbf{R}}) \right] s^{(\alpha\beta\gamma\delta)}(\tilde{\mathbf{r}} - \tilde{\boldsymbol{\tau}}_\kappa - \tilde{\mathbf{R}}) \\ + \rho''_{\kappa c}\left[ s(\tilde{\mathbf{r}} - \tilde{\boldsymbol{\tau}}_\kappa - \tilde{\mathbf{R}}) \right] s^{(\alpha\beta)}(\tilde{\mathbf{r}} - \tilde{\boldsymbol{\tau}}_\kappa - \tilde{\mathbf{R}}) s^{(\gamma\delta)}(\tilde{\mathbf{r}} - \tilde{\boldsymbol{\tau}}_\kappa - \tilde{\mathbf{R}}) \Big\}. \tag{66}$$

The second derivatives of the $xc$ energy are

$$\frac{\partial^2 E_{xc}}{\partial \eta_{\alpha\beta} \partial \eta_{\gamma\delta}} = \delta_{\alpha\beta} \delta_{\gamma\delta} E_{xc}^{(0)} + \Omega \int \left[ \left( K_{xc} n_e^{(0)} - V_{xc}^{(0)} \right) \left( \delta_{\alpha\beta} \delta_{\gamma\delta} n_e^{(0)} - \delta_{\alpha\beta} \frac{\partial n_c}{\partial \eta_{\gamma\delta}} - \delta_{\gamma\delta} \frac{\partial n_c}{\partial \eta_{\alpha\beta}} \right) \right.$$
$$\left. + V_{xc}^{(0)} \frac{\partial^2 n_c}{\partial \eta_{\alpha\beta} \partial \eta_{\gamma\delta}} + K_{xc} \frac{\partial n_c}{\partial \eta_{\alpha\beta}} \frac{\partial n_c}{\partial \eta_{\gamma\delta}} \right] d^3 \tilde{r}. \quad (67)$$

Finally, second-order derivatives with respect to a strain component and a reduced-atomic-displacement component are required. The required "size" derivatives are

$$s^{(i)}(\tilde{\mathbf{r}}) \equiv \frac{\partial s(\tilde{\mathbf{r}})}{\partial \tilde{r}_i} = \sum_j s^{-1}(\tilde{\mathbf{r}}) \Xi_{ij} \tilde{r}_j, \quad (68)$$

and

$$s^{(\alpha\beta i)}(\tilde{\mathbf{r}}) \equiv \frac{\partial^2 s(\tilde{\mathbf{r}})}{\partial \eta_{\alpha\beta} \partial \tilde{r}_i} = \sum_j \left[ s^{-1}(\tilde{\mathbf{r}}) \Xi_{ij}^{(\alpha\beta)} + s^{-3}(\tilde{\mathbf{r}}) s^{(\alpha\beta)}(\tilde{\mathbf{r}}) \Xi_{ij} \right] \tilde{r}_j. \quad (69)$$

The corresponding equations for the $n_c$ derivatives are found by straightforward substitutions $\gamma\delta \to i$ and $\eta_{\gamma\delta} \to \tilde{\tau}_{\kappa i}$ in Eqs.(64) and (66). The xc energy second derivative is

$$\frac{\partial^2 E_{xc}}{\partial \eta_{\alpha\beta} \partial \tilde{\tau}_{\kappa i}} = \Omega \int \left[ \delta_{\alpha\beta} \left( V_{xc}^{(0)} - K_{xc} n_e^{(0)} \right) \frac{\partial n_c}{\partial \tilde{\tau}_{\kappa i}} + V_{xc}^{(0)} \frac{\partial^2 n_c}{\partial \eta_{\alpha\beta} \partial \tilde{\tau}_{\kappa i}} + K_{xc} \frac{\partial n_c}{\partial \eta_{\alpha\beta}} \frac{\partial n_c}{\partial \tilde{\tau}_{\kappa i}} \right] d^3 \tilde{r}. \quad (70)$$

### H. Ion-ion interactions

While not part of $E_{el}$ and not treated in DFPT, the ion-ion interactions are a strain dependent part of the total energy of a solid. They are conventionally evaluated as a sum of three terms using the Ewald summation formula,[32]

$$E_{II}^G = \frac{1}{2\pi\Omega} \sum_{\mathbf{G} \neq 0} \frac{e^{-(\pi G/\xi)^2}}{G^2} \sum_{\kappa\kappa'}^{cell} Z_\kappa Z_{\kappa'} e^{i\mathbf{G}\cdot(\boldsymbol{\tau}_\kappa - \boldsymbol{\tau}_{\kappa'})}, \quad (71)$$

$$E_{II}^R = \frac{1}{2} \sum_{\mathbf{R}} \sum_{\kappa\kappa'}^{cell} Z_\kappa Z_{\kappa'} \frac{\operatorname{erfc}(\xi |\boldsymbol{\tau}_\kappa - \boldsymbol{\tau}_{\kappa'} - \mathbf{R}|)}{|\boldsymbol{\tau}_\kappa - \boldsymbol{\tau}_{\kappa'} - \mathbf{R}|}, \quad (72)$$

$$E_{II}^0 = \frac{-\xi}{\sqrt{\pi}} \sum_\kappa^{cell} Z_\kappa^2, \quad (73)$$

where $Z_\kappa$ are the ion charges and $\xi$ is a convergence parameter. In Eq. (72) and similar equations below, the $\kappa = \kappa'$ term in the sum is to be omitted when $\mathbf{R} = 0$. The strain second derivatives of the reciprocal space sum is similar to Eq. (58), the Hartree term in Sec. IIIE,

$$\frac{\partial^2 E_{II}^G}{\partial \eta_{\alpha\beta} \partial \eta_{\gamma\delta}} = \frac{1}{2\pi\Omega} \sum_{\tilde{\mathbf{G}} \neq 0} \frac{e^{-(\pi G/\xi)^2}}{G^2} \sum_{\kappa\kappa'}^{\text{cell}} Z_\kappa Z_{\kappa'} e^{2\pi i \tilde{\mathbf{G}} \cdot (\tilde{\boldsymbol{\tau}}_\kappa - \tilde{\boldsymbol{\tau}}_{\kappa'})} \Big[ \delta_{\alpha\beta} \delta_{\gamma\delta}$$
$$+ (G^{-2} + \pi^2 \xi^{-2}) \sum_{ij} \Big( \delta_{\alpha\beta} \Upsilon_{ij}^{(\gamma\delta)} + \delta_{\gamma\delta} \Upsilon_{ij}^{(\alpha\beta)} - \Upsilon_{ij}^{(\alpha\beta\gamma\delta)} \Big) \tilde{G}_i \tilde{G}_j \quad (74)$$
$$+ (\pi^4 \xi^{-4} + 2\pi^2 \xi^{-2} G^{-2} + 2 G^{-4}) \sum_{ij} \Upsilon_{ij}^{(\alpha\beta)} \tilde{G}_i \tilde{G}_j \sum_{kl} \Upsilon_{kl}^{(\gamma\delta)} \tilde{G}_k \tilde{G}_l \Big].$$

The strain – reduced-atomic-coordinate second derivative is

$$\frac{\partial^2 E_{II}^G}{\partial \eta_{\alpha\beta} \partial \tilde{\tau}_{\kappa k}} = \frac{i Z_\kappa}{\Omega} \sum_{\tilde{\mathbf{G}} \neq 0} \frac{e^{-(\pi G/\xi)^2}}{G^2} e^{2\pi i \tilde{\mathbf{G}} \cdot \tilde{\boldsymbol{\tau}}_\kappa} \tilde{G}_k \sum_{\kappa'}^{\text{cell}} Z_{\kappa'} e^{-2\pi i \tilde{\mathbf{G}} \cdot \tilde{\boldsymbol{\tau}}_{\kappa'}} \Big[ \delta_{\alpha\beta} + (G^{-2} + \pi^2 \xi^{-2}) \sum_{ij} \Upsilon_{ij}^{(\alpha\beta)} \tilde{G}_i \tilde{G}_j \Big]. \quad (75)$$

The derivatives of the real-space sum involve much of the same analysis as was applied to the model core charge in Sec. IIIG. Let us introduce the compact notation

$$s_{\kappa\kappa'\tilde{\mathbf{R}}} = s\left( \tilde{\boldsymbol{\tau}}_\kappa - \tilde{\boldsymbol{\tau}}_{\kappa'} - \tilde{\mathbf{R}} \right) \quad (76)$$

with a similar subscript notation for the several derivatives of $s$ defined in Eqs.(63), (65), (68), and (69). The strain-strain derivative of the real-space sum is then

$$\frac{\partial^2 E_{II}^R}{\partial \eta_{\alpha\beta} \partial \eta_{\gamma\delta}} = \frac{2}{\sqrt{\pi}} \sum_{\tilde{\mathbf{R}}} \sum_{\kappa\kappa'}^{\text{cell}} Z_\kappa Z_{\kappa'} \Big\{ \Big[ \big( \xi^3 + \xi / s_{\kappa\kappa'\tilde{\mathbf{R}}} \big) e^{-(\xi s_{\kappa\kappa'\tilde{\mathbf{R}}})^2}$$
$$+ \sqrt{\pi} \operatorname{erfc}(\xi s_{\kappa\kappa'\tilde{\mathbf{R}}}) / 2 s_{\kappa\kappa'\tilde{\mathbf{R}}}^3 \Big] s_{\kappa\kappa'\tilde{\mathbf{R}}}^{(\alpha\beta)} s_{\kappa\kappa'\tilde{\mathbf{R}}}^{(\gamma\delta)} \quad (77)$$
$$- \Big[ \xi / 2 s_{\kappa\kappa'\tilde{\mathbf{R}}} + \sqrt{\pi} \operatorname{erfc}(\xi s_{\kappa\kappa'\tilde{\mathbf{R}}}) / 4 s_{\kappa\kappa'\tilde{\mathbf{R}}}^2 \Big] s_{\kappa\kappa'\tilde{\mathbf{R}}}^{(\alpha\beta\gamma\delta)} \Big\}.$$

The corresponding strain – reduced-atomic-coordinate expression is obtained from the analogues of Eqs.(68) and (69), and the substitutions $\gamma\delta \to i$ and $\eta_{\gamma\delta} \to \tilde{\tau}_{\kappa i}$ in Eq. (77).

The Ewald result for the ion-ion interaction represents the energy of an array of point charges interacting with a uniform neutralizing background. In fact, the proper reference is the local pseudopotentials, which differ from Coulombic potentials in their core region, interacting with the uniform background. This energy correction is given by

$$E_{psp-core} = \frac{1}{\Omega} \Big( \sum_{\kappa'} Z_{\kappa'} \Big) \Big( \sum_\kappa 4\pi \int_0^\infty [v_{\kappa Loc}(r) + Z_\kappa / r] r^2 dr \Big). \quad (78)$$

The only strain dependence is through the $\Omega^{-1}$ factor, so the second strain derivative of this term is simply $\delta_{\alpha\beta} \delta_{\gamma\delta} E_{psp-core}$.

## IV. IMPLEMENTATION AND RESULTS

### A. Clamped-atom perturbations

The metric tensor formulation of strain perturbations in DFPT was developed and tested in stages within the open-source ABINIT software package.[33] As anticipated, it could be merged cleanly into the existing DFPT structure of this code which had previously been developed to treat atomic-displacement and electric-field perturbations. The ground-state portions of this code already calculated relevant first derivatives of the DFT total energy, in particular atomic forces, stresses using the Nielsen-Martin analysis,[1] and polarization using the Berry-phase method.[7,17] The availability of first derivatives calculated in a context completely consistent with the newly developed strain second derivatives permitted critical comparisons to verify the new formalism and its computational realization.

Numerical strain derivatives of the various first derivatives were carried out using the 5-point formula,[34] and strain increments sufficiently small to ensure an invariant set of $\tilde{\mathbf{K}}$ within the specified energy cutoffs. These comparisons required consistency between the ground-state DFT and DFPT calculations with regard to cutoffs, Brillouin zone sampling, etc., but not necessarily complete convergence with respect to these parameters. What was required for accurate comparisons was an exceedingly high level of convergence of the self-consistent potentials and wave functions, both for the ground-state numerical derivatives and the DFPT results. This was necessitated by the fact that the expressions used for the mixed second derivatives, Eqs.(15) and (16), are non-stationary, and such convergence errors appear in first-order.

The level of agreement that can be obtained for the elastic and piezoelectric tensors is illustrated in Tables I and II, respectively. The system chosen for this example was AlP, but with the two-atom unit cell of the zincblende structure randomly distorted in the range ±5% for both the primitive lattice vectors and the relative atomic positions. This was necessary to obtain a full set of tensor elements for comparison, since most would otherwise be zero or identical because of symmetry. Stresses in the reference (nominally "unstrained") configuration were not relaxed, so the elastic tensor second-derivatives needed to be compared to $\Omega^{-1}\partial(\Omega\sigma_{\alpha\beta})/\partial\eta_{\gamma\delta}$ rather than $\partial\sigma_{\alpha\beta}/\partial\eta_{\gamma\delta}$, following Eq. (32). This also enhanced the completeness of these tests, since a subset of the terms derived in Sec. III would mutually cancel for a truly unstrained reference structure.

For the piezoelectric tensor comparisons, there are two caveats. Eq. (16) requires first-order wave functions $\psi_\alpha^{\tilde{k}_j}$ for the $d/dk$ perturbation, which are best found from DFPT.[9] However, the ground-state calculations of the polarization perform Berry-phase integrations on a discrete grid of **k** points in the Brillouin zone.[7,17] For optimum consistency, a finite-difference approximation to the $\psi_\alpha^{\tilde{k}_j}$ based on the ground-state grid was used.[35] In the limit of a large **k** sample, both approaches give the same result as they must. Results with the DFPT $\psi_\alpha^{\tilde{k}_j}$ in fact converge much more rapidly with zone sample size. The second issue concerns the effects on the strain numerical derivatives of the polarization of the reference configuration. The straight numerical derivatives yield the so-called "improper" piezoelectric tensor, while DFPT yields the "proper" tensor. Knowing the reference configuration polarization, the proper tensor can be calculated from the im-proper one in a straightforward manner,[36] and this has been done for the comparisons in Table II.

**B. Relaxed-atom calculations**

While homogeneous strain as defined in Eq. (17) moves all atoms proportionally, in a real experimental situation macroscopic strain only deforms the unit cells, and the atomic positions readjust. The effects of this relaxation on the elastic and piezoelectric tensors can be calculated analytically as corrections to the clamped-atom quantities. These corrections can be computed from the set of mixed second derivatives with respect to one strain component and one component of each internal atomic coordinate, the "internal strain."[37] The expressions needed to compute the frozen-wave-function contributions to internal strain have been given in Sec. III for each term in the DFT energy. We have used the non-stationary expression for mixed second derivatives, Eq. (15), with the strain-perturbation wave function for $\psi^{(\lambda_2)}$ and the atomic-coordinate component first-order Hamiltonian for $H^{(\lambda_1)}$, whose terms have been given previously.[9]

The relaxation corrections also require mixed second derivatives with respect to pairs of internal-atomic-coordinate components, known as the interatomic force constant matrix, and with respect to one atomic-coordinate component and one electric-field component, known as the Born effective charges.[37] The DFPT expressions needed to evaluate these quantities have also been given,[9] and were previously implemented in the ABINIT package.[33] The expressions combining all these mixed derivatives to obtain the atomic-relaxation corrections are straightforward, and will not be detailed here.[37]

Numerical-derivative comparisons including the relaxations are especially challenging. In addition to the considerations discussed above for consistency and convergence of the clamped-atom quantities, the atomic positions in the incrementally-strained unit cells must be relaxed in the ground-state DFT calculations until the forces are far smaller than typically considered necessary for structural optimization. Tables III and IV give the relaxed-atom results for the elastic and piezoelectric tensors for the distorted AlP example discussed above. The agreement between the numerical derivatives and the DPFT results are excellent, but respectively one and two orders of magnitude worse on the average than for the clamped-atom quantities. This level of agreement required attaining residual forces less than $10^{-10}$ atomic units (Hartree/Bohr) for the $2\times10^{-5}$ strain increment needed to satisfy the conditions discussed above. The precision of the required relaxation illustrates the impracticality of obtaining accurate values for the relaxed-atom quantities for more complex systems by numerical differentiation. Attempts at further convergence suggested that the level of agreement shown here is at the limit of numerical precision for the overall set of calculations.

Comparing the tables of relaxed and unrelaxed tensors, we see that the relaxation corrections to the large components of the elastic tensor, those which would be present for the zincblende structure without the random distortions, are rather small. For the piezoelectric tensor however, the only large component, (x, yz), is substantially corrected.

## V. SUMMARY AND CONCLUSIONS

In conclusion, we have demonstrated the manner in which strain can be treated within a standard implementation of density-functional perturbation theory by using reduced coordinates and the subsequent strain dependence of the metric tensor. Expressions necessary to evaluate all the second-order derivatives of the density functional theory energy have been derived, and it has been established that they are correct and complete by comparisons with numerical derivatives. Direct calculation of the elastic and piezoelectric tensors, including atomic relaxation, is thereby achieved. The level of agreement with experimental quantities is, of course, determined by the fundamental limitations of density-functional theory, and to a lesser extent by the pseudopotential approximation and the quality of the pseudopotentials which are employed.

The expressions given here pertain to norm-conserving pseueopotentials.[11] While the same approach can in principle be applied to ultrasoft pseudopotentials,[38] the closely related projector-augmented-wave all-electron method,[39] and the linear-augmented-plane-wave method,[40] these all pose significant additional challenges. The first set of challenges relates to the fact that the nonlocal operators coupling the plane-wave components of these methods have off-diagonal terms compling the $\ell m$, $\ell' m'$ spherical harmonic indices about each atomic site. This precludes the reduction to wave-vector dot products achieved in Eq. (49). The second issue concerns the augmentation components of the wave functions and charge. These functions are not deformed by homogeneous strain in the manner of the plane waves and plane-wave charge density. Thus the mapping onto reduced coordinates and derivatives of that mapping entail issues similar to those discussed in Sec. III G in connection with model core charges and the nonlinear core correction.[31] Unlike the core charges, however, the augmentation function are not spherical, so additional considerations apply. While the implementation of the strain perturbation within DFPT using these formalisms poses these challenges and requires significant further analysis, the metric tensor approach likely remains the most viable.

**Acknowledgements**

We would like to thank X. Gonze, D. C. Allan, and A. R. Oganov for valuable discussions of aspects of this work, and for access to their unpublished notes. This work was supported in part by the Center for Piezoelectrics by Design, ONR grant N00014-01-1-0365.

Table I. Comparison of a sample of numerical and DFPT clamped-atom elastic tensor components (GPa) for distorted AlP. The strain increment for numerical differentiation is $2\times10^{-5}$. The overall rms difference is $5\times10^{-6}$ GPa.

|    |    | Numerical      | DFPT           | Difference  |
|----|----|----------------|----------------|-------------|
| xx | xx | 1.2499990E+02  | 1.2499990E+02  | 2.400E-05   |
| yy | xx | 6.6990360E+01  | 6.6990360E+01  | -3.900E-06  |
| zz | xx | 6.8396840E+01  | 6.8396840E+01  | -1.500E-06  |
| yz | xx | 8.8373390E-02  | 8.8373500E-02  | 1.054E-07   |
| xz | xx | -1.1173330E+00 | -1.1173330E+00 | -4.300E-07  |
| xy | xx | -4.1892180E-01 | -4.1892170E-01 | 5.700E-08   |
|    |    |                |                |             |
| xx | yz | 8.8374160E-02  | 8.8373500E-02  | -6.607E-07  |
| yy | yz | 5.1544700E+00  | 5.1544690E+00  | -1.030E-06  |
| zz | yz | -5.5782700E+00 | -5.5782700E+00 | -2.900E-07  |
| yz | yz | 9.0315730E+01  | 9.0315730E+01  | 4.500E-06   |
| xz | yz | -4.0474890E-01 | -4.0474890E-01 | 5.000E-08   |
| xy | yz | 6.4472760E-01  | 6.4472770E-01  | 6.400E-08   |

Table II. Comparison of a sample of numerical and DFPT clamped-atom piezoelectric tensor components (C/m$^2$) for distorted AlP. The strain increment for numerical differentiation is $2\times10^{-5}$. The overall rms difference is $2\times10^{-8}$ C/m$^2$.

|   |    | Numerical      | DFPT           | Difference |
|---|----|----------------|----------------|------------|
| x | xx | 2.0211410E-02  | 2.0211400E-02  | -8.700E-09 |
| y | xx | 5.2336140E-02  | 5.2336120E-02  | -1.770E-08 |
| z | xx | 4.0031790E-03  | 4.0031860E-03  | 6.720E-09  |
|   |    |                |                |            |
| x | yy | -8.2697310E-02 | -8.2697310E-02 | 3.000E-10  |
| y | yy | 2.4712180E-03  | 2.4712150E-03  | -3.280E-09 |
| z | yy | 7.3837080E-03  | 7.3837040E-03  | -4.260E-09 |
|   |    |                |                |            |
| x | yz | -6.9263100E-01 | -6.9263100E-01 | -3.600E-08 |
| y | yz | -1.4235180E-03 | -1.4235300E-03 | -1.231E-08 |
| z | yz | -1.3531730E-02 | -1.3531760E-02 | -2.880E-08 |

Table III. Comparison of a sample of numerical and DFPT relaxed-atom elastic tensor components (GPa) for distorted AlP. The strain increment for numerical differentiation is $2\times10^{-5}$. The overall rms difference is $4\times10^{-5}$ GPa.

|  |  | Numerical | DFPT | Difference |
|---|---|---|---|---|
| xx | xx | 1.2499150E+02 | 1.2499150E+02 | -1.100E-05 |
| yy | xx | 6.6999750E+01 | 6.6999760E+01 | 8.200E-06 |
| zz | xx | 6.8359440E+01 | 6.8359440E+01 | 7.000E-07 |
| yz | xx | 2.2844680E -01 | 2.2846610E -01 | 1.927E-05 |
| xz | xx | -1.1398380E -00 | -1.1398280E -00 | 9.560E-06 |
| xy | xx | -1.5027680E -02 | -1.5117250E -02 | -8.957E-05 |
|  |  |  |  |  |
| xx | yz | 2.2847050E -01 | 2.2846610E -01 | -4.380E-06 |
| yy | yz | 1.9400500E -00 | 1.9400540E -00 | 3.720E-06 |
| zz | yz | -2.0792640E -00 | -2.0792750E -00 | -1.109E-05 |
| yz | yz | 6.6593340E+01 | 6.6593390E+01 | 5.160E-05 |
| xz | yz | 7.7397220E -01 | 7.7397730E -01 | 5.121E-06 |
| xy | yz | -5.6844590E-01 | -5.6844910E -01 | -3.170E-06 |

Table IV. Comparison of a sample of numerical and DFPT relaxed-atom piezoelectric tensor components (C/m$^2$) for distorted AlP. The strain increment for numerical differentiation is $2\times10^{-5}$. The overall rms difference is $2\times10^{-6}$ C/m$^2$.

|  |  | Numerical | DFPT | Difference |
|---|---|---|---|---|
| x | xx | 1.7147690E-02 | 1.7146940E-02 | -7.461E-07 |
| y | xx | 5.1070690E-02 | 5.1070800E-02 | 1.069E-07 |
| z | xx | -8.8396190E-03 | -8.8367620E-03 | 2.857E-06 |
|  |  |  |  |  |
| x | yy | 8.2856910E-03 | 8.2845410E-03 | -1.150E-06 |
| y | yy | 3.7168430E-02 | 3.7168120E-02 | -3.150E-07 |
| z | yy | -8.1020100E-03 | -8.1017610E-03 | 2.494E-07 |
|  |  |  |  |  |
| x | yz | -3.8719800E-02 | -3.8721540E-02 | -1.739E-06 |
| y | yz | -1.2451730E-02 | -1.2452060E-02 | -3.271E-07 |
| z | yz | 1.9026870E-02 | 1.9026930E-02 | 5.590E-08 |